\documentclass[conference]{IEEEtran}
\usepackage{amsmath,amssymb,amsfonts}
\usepackage{algorithmic}
\usepackage{graphicx}
\usepackage{textcomp}
\usepackage{xcolor}
\usepackage{balance}
\usepackage{tabularx}  % For equal column width
\usepackage{booktabs}  % For better looking tables
\usepackage{fancyhdr}
\usepackage{url}
\usepackage[square,comma,sort&compress,numbers]{natbib} 
\usepackage{threeparttable,multirow,adjustbox,threeparttable}
\usepackage[symbol]{footmisc}
\usepackage{orcidlink} % orcidlink
\usepackage{adjustbox}% 引入超链接宏包
\usepackage{rotating} % 引入rotating包
\usepackage{verbatim}
\usepackage{pbalance}  % 最后一页平衡
\usepackage{hyperref}
\hypersetup{
    colorlinks=true,        % 激活链接颜色，去掉链接边框
    linkcolor=blue,         % 文档内部链接颜色（如图表等引用）
    citecolor=blue,         % 文献引用链接颜色
    filecolor=blue,         % 文件链接颜色
    urlcolor=blue           % 外部URL链接颜色
}
\usepackage{caption}
\begin{document}
\title{Masked Graph Autoencoders with Contrastive Augmentation for Spatially Resolved Transcriptomics Data} 
% 一种基于图神经网络的空间转录组多切片空间域识别及批次效应去除方法
%-------------------------------------------------------------------------------------
\author{
\IEEEauthorblockN{
Donghai Fang$^{1,\dag}$, 
Fangfang Zhu$^{2,\dag}$, 
Dongting Xie$^1$ and 
Wenwen Min$^{1*}\orcidlink{0000-0002-2558-2911}$  }\\
\IEEEauthorblockA{
$^1$School of Information Science and Engineering, Yunnan University, Kunming 650091, Yunnan, China \\
$^2$College of Nursing Health Sciences, Yunnan Open University,  Kunming 650599, Yunnan, China\\
$\dag$Co-first authors and $^{*}$Correspondence author: minwenwen@ynu.edu.cn.}				
}
\maketitle
\thispagestyle{fancy}
\begin{abstract}
With the rapid advancement of Spatial Resolved Transcriptomics (SRT) technology, it is now possible to comprehensively measure gene transcription while preserving the spatial context of tissues. Spatial domain identification and gene denoising are key objectives in SRT data analysis. We propose a Contrastively Augmented Masked Graph Autoencoder (STMGAC) to learn low-dimensional latent representations for domain identification. In the latent space, persistent signals for representations are obtained through self-distillation to guide self-supervised matching. At the same time, positive and negative anchor pairs are constructed using triplet learning to augment the discriminative ability. We evaluated the performance of STMGAC on five datasets, achieving results superior to those of existing baseline methods. All code and public datasets used in this paper are available at \url{https://github.com/wenwenmin/STMGAC} and \url{https://zenodo.org/records/13253801}.
\end{abstract}


\section{Introduction}
In complex organisms, cells form functionally specialized clusters through dynamic interactions and intricate organizational structures \cite{ncbenchmark}. These clusters coordinate the functions of the organism through mutual influences and tight connections. The latest spatial resolved transcriptomics (SRT) technologies, such as ST, 10x Visium, and Stereo-seq, can comprehensively measure transcriptional expression at specific spatial locations (spots) while preserving the spatial context of the tissue \cite{STAGATE}.

In the analysis of SRT data, the key computational tasks are identifying shared and specific clusters \cite{BIBM1}, known as spatial domains, and addressing data denoising issues \cite{li2024stmcdi}. Traditional non-spatial clustering methods, such as the Louvain algorithm \cite{louvain}, have failed to effectively utilize spatial information, resulting in incoherent clustering results within tissue sections.

Recently, proposed spatial clustering methods consider the similarity between neighboring points to reveal the spatial dependence of gene expression. For example, DeepST \cite{DeepST} uses denoising autoencoders and graph neural network (GNN) autoencoders to jointly infer latent embeddings of enhanced spatial transcriptomics data. SEDR \cite{SEDR} employs variational graph autoencoders based on masking mechanisms to infer latent representations. STAGATE \cite{STAGATE} integrates spatial information and gene expression through an adaptive graph autoencoder to learn low-dimensional representations. GraphST \cite{graphst} adopts a graph self-supervised contrastive learning framework to learn the latent embeddings of spatial transcriptomics data. DiffusionST \cite{Diffusionst} uses a zero-inflated negative binomial (ZINB) distribution and a diffusion model for data denoising and enhancement. Although these methods aim to reduce redundancy in transcriptional expression and employ various techniques to address this issue, they overlook the supervisory signals in the latent space \cite{min2024dimensionality, graphmae, AFGRL}. Furthermore, the complex graph self-supervised contrastive learning frameworks lack consideration of global semantic information and the similarity of spatial structures \cite{RARE, zhiceng}. 

To address the aforementioned challenges, we propose a Contrastively Augmented Masked Graph Autoencoder (STMGAC) to learning low-dimensional latent representations for SRT data analysis. First, we use a masked graph autoencoder to reconstruct raw gene expression and perform gene denoising. Next, we obtain persistent and reliable latent space supervision information through self-distillation, which guides the latent representation for self-supervised matching and results in more stable low-dimensional embeddings. Additionally, we use a plug-and-play method for selecting positive and negative anchor pairs in the latent space, leveraging triplet learning to reduce the distance between embeddings of similar items while ensuring better separation of items from different categories, which improves spatial domain clustering. To validate the effectiveness of STMGAC, we compared it with seven advanced methods across five SRT datasets from three different platforms and conducted extensive ablation studies to investigate the contributions of various components. Experimental results show that STMGAC outperforms other state-of-the-art methods in spatial clustering and trajectory inference.

The main contributions of our proposed method are:
\begin{itemize}
    
    \item We propose a Contrastively Augmented Masked Autoencoder (STMGAC) to learn low-dimensional latent representations for SRT data analysis. 

    \item Using self-distillation learning, we acquire latent space supervision signals that improve the reconstruction of raw gene expressions.

    \item We employ positive and negative anchor pairs for triplet learning to achieve superior spatial clustering.

    \item Our method achieves superior accuracy metrics compared to existing state-of-the-art methods across five datasets.
\end{itemize}

%%%%%%%%%%%%%%%%%%%%%%%%
\begin{figure*}[t]%
\centering
\centering
\includegraphics[width=1\textwidth]{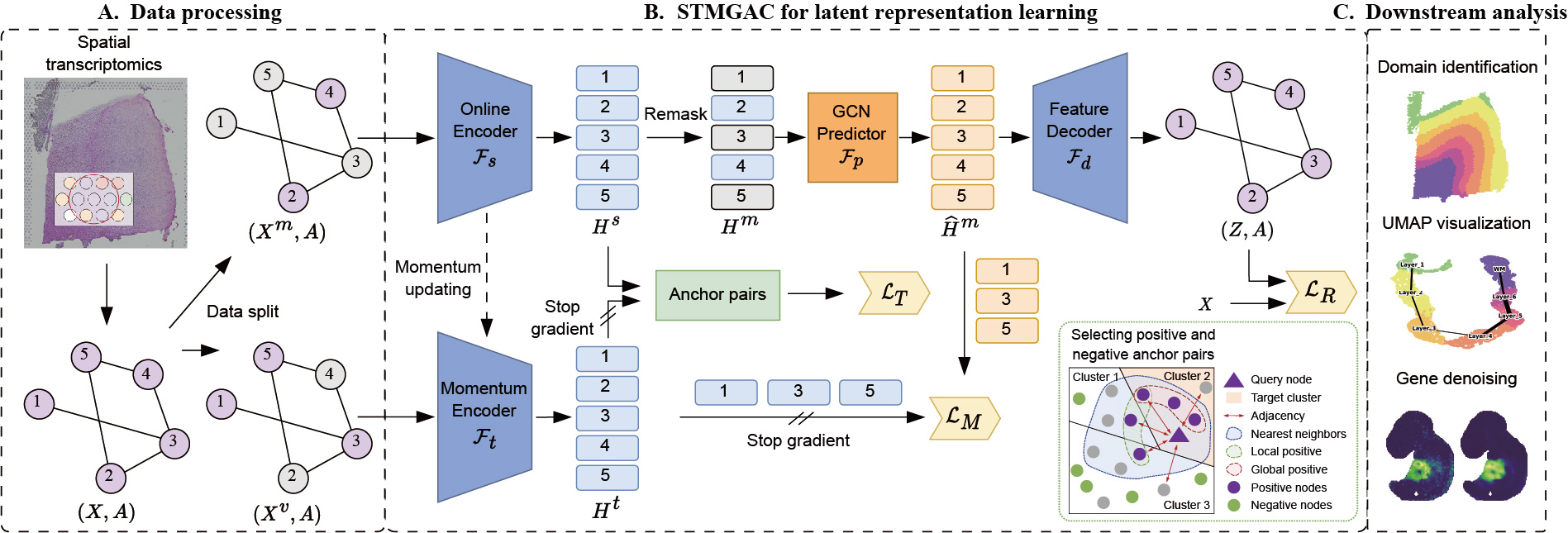}
\caption{Overview of STMGAC. (A) In data processing, raw gene expression is split into a mask matrix and a visible matrix. (B) Reconstruct raw data using a mask GAE, obtain latent space supervision signals through a momentum encoder for latent representation matching, and then utilize selected anchor spots for triplet learning. (C) The learned latent representations from STMGAC will be utilized in downstream task analysis, including clustering and visualization. Additionally, the reconstructed gene expression is considered as the denoised outcome.}
\label{fig1}
\end{figure*}

%%%%%%%%%%%%%%%%%%%%%%%%
\section{PROPOSED METHODS}\label{MATERIALS AND MRTHODS}
\subsection{Overview of the proposed STMGAC}
In data processing, the raw gene expression data is divided into a mask matrix and a visible matrix. A mask GAE is used to reconstruct the raw data, and latent space supervision signals are obtained through a momentum encoder for matching latent representations. Selected anchor spots are then employed for triplet learning. The learned latent representations from STMGAC will be used in downstream analyses (\autoref{fig1}).

\subsection{Data preprocessing and spatial graph construction}
STMGAC uses gene transcription expression from SRT data and the spatial coordinates of spots as input. Using the library functions provided by SCANPY to retain the genes of interest, and then log-normalize the entire expression. Finally, retain the top $N_g$ highly variable genes to obtain the preprocessed data $X\in \mathrm{R}^{N\times N_g}$, where $N$ is the total number of spots. 

Calculate the Euclidean distance between spots based on their spatial coordinates, and then apply the $K$-nearest neighbors (KNN) algorithm to select the $K$ nearest spots, thereby constructing the adjacency matrix $A$. If spot $j$ is a neighbor of spot $i$, then $A_{ij} = A_{ji} = 1$. The constructed adjacency matrix will be used in each involving graph neural network.

\subsection{Data augmentation with spot masking}
Before training STMGAC, we first generate two complementary masked gene expression graphs. They will be used as the input for the online encoder and momentum encoder, respectively. Specifically, with a masking rate $\rho$, we randomly sample a masked subset $\mathcal{V}^m$ from the set of all spots $\mathcal{V}$ in the SRT data, while the remaining spot set forms the visible subset $\mathcal{V}^v$, satisfying $\mathcal{V}^m\cup \mathcal{V}^v=\mathcal{V}$ and $\mathcal{V}^m\cap \mathcal{V}^v=\varnothing$. 

To overcome the ``identity mapping" problem (i.e., directly mapping the input to the output) and to reduce redundancy in SRT data, we construct the masked matrix $X^m \in \mathrm{R}^{N\times N_g}$ of the gene expression matrix. It is defined as follows: for any spot $i$ ($v_i$), if $v_i\in \mathcal{V}^m$, then its corresponding gene expression value is replaced with a learnable mask token $x_{[M]} \in \mathrm{R}^{N_g}$, i.e., $x^m_i=x_{[M]}$; otherwise, $x^m_i=x_i$. 

To ensure that the spots masked in the latent space have persistent supervision signals and perform latent feature matching, we construct the complementary visible matrix $X^v \in \mathrm{R}^{N\times N_g}$ of $X^m$. It is defined as follows: if $v_i\in \mathcal{V}^v$, then its corresponding gene expression value is replaced with the mask token, i.e., $x^v_i=x_{[M]}$; otherwise, $x^v_i=x_i$.

\subsection{Latent representation learning with masked reconstruction}
\subsubsection{Graph encoding} The online encoder $\mathcal{F}_s$ is responsible for converting the masked matrix $X^m$ into a low-dimensional latent representation. To achieve this, we first use a MLP $\mathcal{F}_{s,f}$ composed of stacked linear layers for initial dimensionality reduction to obtain the low-dimensional feature representation $H^f \in \mathrm{R}^{N\times d_f}$. Here, $d_f$ is the dimensions of the feature.
\begin{equation}
\small{
    H^f=\mathcal{F}_{s,f}(X^m;\Theta_{s,f})=W^1_f\text{ELU}(\text{BN}(W^0_fX^m+b^0_f))+b^1_f}
\end{equation}
where $W_f^l$ and $b_f^l$ represent the weights and biases of the $l$-th layer of the linear layer. BN is batch normalization. 

Then, a graph encoder $\mathcal{F}_{s,g}$ consisting of two layers of GCN forces the model to learn an effective latent representation $H^s \in \mathrm{R}^{N\times d}$ from the visible spot neighbors. Here, $d$ is the dimensions of the latent representation. This process is represented as follows: 
\begin{equation}
    \small{H^s=\mathcal{F}_{s,g}(H^f,A;\Theta_{s,g})=\widetilde{A}\text{ReLU}(\text{BN}(\widetilde{A}H^fW^0_g))W^1_g}
\end{equation}
where $W_g^l$ represents the weights of the $l$-th layer of the GCN, as well as the symmetrically normalized adjacency matrix $\widetilde{A}=D^{-\frac{1}{2}}AD^{-\frac{1}{2}}$. Therefore, the calculation of $H^s$ can be expressed as: 
\begin{equation}
    \small{H^s=\mathcal{F}_s(X^m,A;\Theta_{s})=\mathcal{F}_{s,g}(\mathcal{F}_{s,f}(X^m;\Theta_{s,f}),A;\Theta_{s,g})}
\end{equation}
where $\Theta_{s}$ is the parameter of the online encoder $\mathcal{F}_s$.

Once the training phase is completed, we utilize X as the input to the online encoder $\mathcal{F}_s$ and obtain the latent representation $H$. This representation is then used for downstream tasks such as spatial domain identification and visualization. 

\subsubsection{Latent representation predicting} The representation predictor $\mathcal{F}_p$ can obtain the predicted expression of the latent representation. To alleviate redundancy, we use the remask technique to mask the representation of the spots in the masked subset $\mathcal{V}^m$ in the latent space, obtaining the remasked representation $H^m \in \mathrm{R}^{N\times d}$. Specifically, for any $v_i$, if $v_i\in \mathcal{V}^m$, then its corresponding latent feature is replaced with a learnable mask token $h_{[RM]}\in \mathrm{R}^{d}$, i.e., $h^m_i=h_{[RM]}$; otherwise, $h^m_i=h^s_i$. 

Unlike typical self-supervised encoders, our latent representation undergoes the remask technique before being input to the predictor. Therefore, we need to use a graph neural network to allow the masked spots to learn the current features from the visible spots. The final predicted latent feature representation is $\widehat{H}^m\in \mathrm{R}^{N\times d}$, calculated as follows: 
\begin{equation}\small{
    \widehat{H}^m=\mathcal{F}_p(H^m,A;\Theta_p)=W^1_p\text{ReLU}(\text{BN}(\widetilde{A}H^mW^0_p))}
\end{equation}
where $W^0_p$ and $W^1_p$ are the weight matrices, and $\Theta_{p}$ is the parameter of the predictor $\mathcal{F}_p$.

The obtained predicted representation matrix $\widehat{H}^m$ will be used to reconstruct the transcription expression of the raw data space and to align with the supervision signal obtained by the momentum encoder.

\subsubsection{Feature decoding} Using an asymmetric autoencoder, a single-layer GCN is used as the decoder $\mathcal{F}_d$ to map $\widehat{H}^m$ into the raw data space, obtaining the reconstructed gene expression matrix $Z \in \mathrm{R}^{N\times N_g}$. The calculation is as follows: 
\begin{equation}\small{Z=\mathcal{F}_d(\widehat{H}^m,A;\Theta_d)=\widetilde{A}\widehat{H}^mW_d}
\end{equation}
where $W_d$ is the weight matrix, and $\Theta_{d}$ is the parameter of the decoder $\mathcal{F}_d$.
\subsubsection{Reconstruction loss in the raw data space} One of the main objectives of STMGAC is to reconstruct the masked gene expression of spots in $\mathcal{V}_m$ given a partially observed spot set and adjacency relationships. By utilizing the Scaled Cosine Loss (SCE) as the objective function, it is defined under a predetermined scaling factor $\gamma$ as follows:
\begin{equation}
    \mathcal{L}_{R}=\frac{1}{|\mathcal{V}_m|}\sum_{v_i\in\mathcal{V}_m}{(1-\frac{x_iz_i^{\mathsf{T}}}{\left\|x_i\right\|\left\|z_i\right\|})^{\gamma}} ,\gamma\ge 1
\end{equation}
where $\gamma$ is fixed to 2 to reduce the contribution from simple samples during the training process, and $|\mathcal{V}_m|$ represents the number of spots in the masked set. 

\subsection{Latent representation matching for robustness enhancement}
\subsubsection{Momentum graph encoding} In graphs, masked spots find it difficult to recover the raw semantics only from their neighbors due to a lack of stronger supervision signals. Therefore, we need to provide each masked spot with a persistent signal in the latent space to guide its self-supervised matching, and this signal is obtained from the data itself through self-distillation learning. 

Specifically, we have a momentum graph encoder $\mathcal{F}_t$ with the same architecture as the graph encoder $\mathcal{F}_s$, responsible for encoding the raw gene expression data $X^v$ in the masked visible set $\mathcal{V}^v$ to obtain a persistent momentum latent representation $H^t\in \mathrm{R}^{N\times d}$ that provides stable guidance for the latent representation $H^s$. The calculation is as follows:
\begin{equation}
    \small{H^t=\mathcal{F}_t(X^v,A;\Theta_{t})}
\end{equation}
where $\Theta_{t}$ is the parameter of the encoder $\mathcal{F}_t$. It is worth noting that $\mathcal{F}_t$ is detached from the gradient back-propagation, and its parameters are updated by exponential moving average (EMA), with the update process as follows:
\begin{equation}
    \small{
    \Theta_{t}\gets \mu \Theta_{t} + (1-\mu)\Theta_{s}
    }
\end{equation}
where $\mu$ is the smoothing factor, fixed at 0.98 based on experimental experience, retaining reliable information for about 50 steps.

\subsubsection{Matching loss in the latent representation space} Here we focus more on feature-level similarity, emphasizing that the predicted latent representation $\widehat{H}^m$ should precisely match the latent representation $H^t$ calculated by the momentum graph encoder. To this end, we use matching loss to minimize their distance:
\begin{equation}\small{\mathcal{L}_M=\frac{1}{|\mathcal{V}_m|}\sum_{v_j\in\mathcal{V}_m}\left\| \widehat{h}^m_j-h^t_j  \right\|^2
    }
\end{equation}

\subsection{Spot triplet learning with positive and negative anchors}
\subsubsection{Selecting positive and negative anchor pairs} Once the low-dimensional representation can reconstruct the high-dimensional raw gene expression, it indicates that the low-dimensional latent representation contains rich semantic information. Therefore, we define triplets in the latent space. Inspired by AFGRL \cite{AFGRL}, we can utilize the encoding of the online and momentum encoders based on their global semantic information and local neighboring structure information to select positive anchors. Unlike previous methods, we further construct negative anchors to form anchor pairs, and experiments have shown that this is more suitable for spatial domain clustering analysis of SRT datasets.

Specifically, for a given query spot \(v_i \in \mathcal{V}\), we calculate the cosine similarity between \(h^s_i\) and all other spots \(h^t_j\) as follows: 

\begin{equation} \small{
    \text{sim}(v_i, v_j) = \frac{h^s_i \cdot h^t_j}{\left\| h^s_i \vphantom{h^t_j} \right\| \left\| h^t_j \vphantom{h^s_i} \right\|}, \forall v_j \in \mathcal{V}}
\end{equation}

After obtaining the similarity information, we calculate the $k$-nearest neighbor set \(\mathcal{B}_i\) in the latent space for each spot \(i\) and use it as a reasonable positive anchor candidate for spot \(v_i\). However, this approach neglects the local structural information between neighboring spots and the global semantics of spots that might belong to the same spatial domain. Therefore, we have the local positive anchor set \(\mathcal{L}_i\) and the global positive anchor set \(\mathcal{G}_i\), satisfying \(\mathcal{L}_i = \mathcal{B}_i \cap \mathcal{A}_i\), where \(\mathcal{A}_i\) represents the neighboring information of the spatial location of the \(i\)-th spot; and \(\mathcal{G}_i = \mathcal{B}_i \cap \mathcal{C}_i\), where \(\mathcal{C}_i\) is the global semantic spot set that belongs to the same spatial domain as the \(i\)-th spot obtained using a clustering algorithm. Therefore, we provide the real positive set \(\mathcal{P}_i = \mathcal{L}_i \cup \mathcal{G}_i\) for spot \(v_i\), considering both local and global information.

We define the negative candidate set for the query spot \( v_i \) as \( {\mathcal{N}}'_i = \mathcal{V} \setminus (\mathcal{B}_i \cup \mathcal{C}_i \cup \mathcal{A}_i) \). The negative set is then obtained by randomly selecting \( n = |\mathcal{P}_i| \) elements from \( {\mathcal{N}}'_i \). Formally, let \(\mathcal{N}_i\) denote this subset, which can be expressed as: \(\mathcal{N}_i \subseteq {\mathcal{N}}'_i\) and \(|\mathcal{N}_i| = n\). Finally, we obtain the anchor pairs \(\mathcal{T}_i = (\mathcal{P}_i, \mathcal{N}_i)\) for the query spot \(v_i\).

\subsubsection{Triplet loss enhances discriminative ability} Triplet loss encourages similar instances (anchors and positive samples) to be closer in the latent space, while dissimilar instances (anchors and negative samples) are pushed further apart. This tightens the clustering of similar items and clearly separates different items, which significantly improves spatial domain recognition capabilities. The calculation is as follows:
\begin{equation}
\small{
    \mathcal{L}_T=\frac{1}{N_t}\sum_{v_i\in\mathcal{V}}\sum_{(p_i, n_i)\in\mathcal{T}_i} \max(\left \| h_i^s-h_p^s \right \|_2 -\left \| h_i^s-h_n^s \right \|_2+ \tau,0) }
\end{equation}
where \(N_t\) represents the total number of anchor pairs, and \(\tau\) is the margin (default 1.0).

Finally, the entire learning objective is written as:
% \begin{equation}\small{
%     \mathcal{L}oss=\frac{\lambda_1\lambda_2}{1-\lambda_2} \mathcal{L}_R+\frac{(1-\lambda_1)\lambda_2}{1-\lambda_2}\mathcal{L}_M+\mathcal{L}_T
%     }
% \end{equation}

\begin{equation}\small{
    \mathcal{L}oss=\lambda_0\mathcal{L}_R+(1-\lambda_0)\mathcal{L}_M+\lambda_1\mathcal{L}_T
    }
\end{equation}
where \(\lambda_0, \lambda_1 \) are hyperparameters that control the contributions to the loss function.

\subsection{Evaluation criteria}
We utilize accuracy metrics to describe the clustering precision of the method. Specifically,  we have the following metrics: Adjusted Rand Index (ARI), used to compare the similarity between clustering results and manually annotated labels. Normalized Mutual Information (NMI), based on information theory, it measures the normalized mutual information between clustering results and true labels. Homogeneity (HOM) score, a metric that assesses if all clusters contain only data points belonging to a single class, indicating homogeneous clustering. Completeness (COM) score, a metric that measures if all data points belonging to a particular class are grouped together in the same cluster, indicating complete clustering \cite{evaluation}. Therefore, the overall accuracy score is calculated as follows:
\begin{equation}\small{
    \text{ACC} = \frac{1}{3} \times (\text{NMI} + \text{HOM} + \text{COM})}
\end{equation}
The closer the ARI and ACC scores are to 1, the better the clustering precision.

\begin{figure*}[!t]%
\centering
\includegraphics[width=1\textwidth]{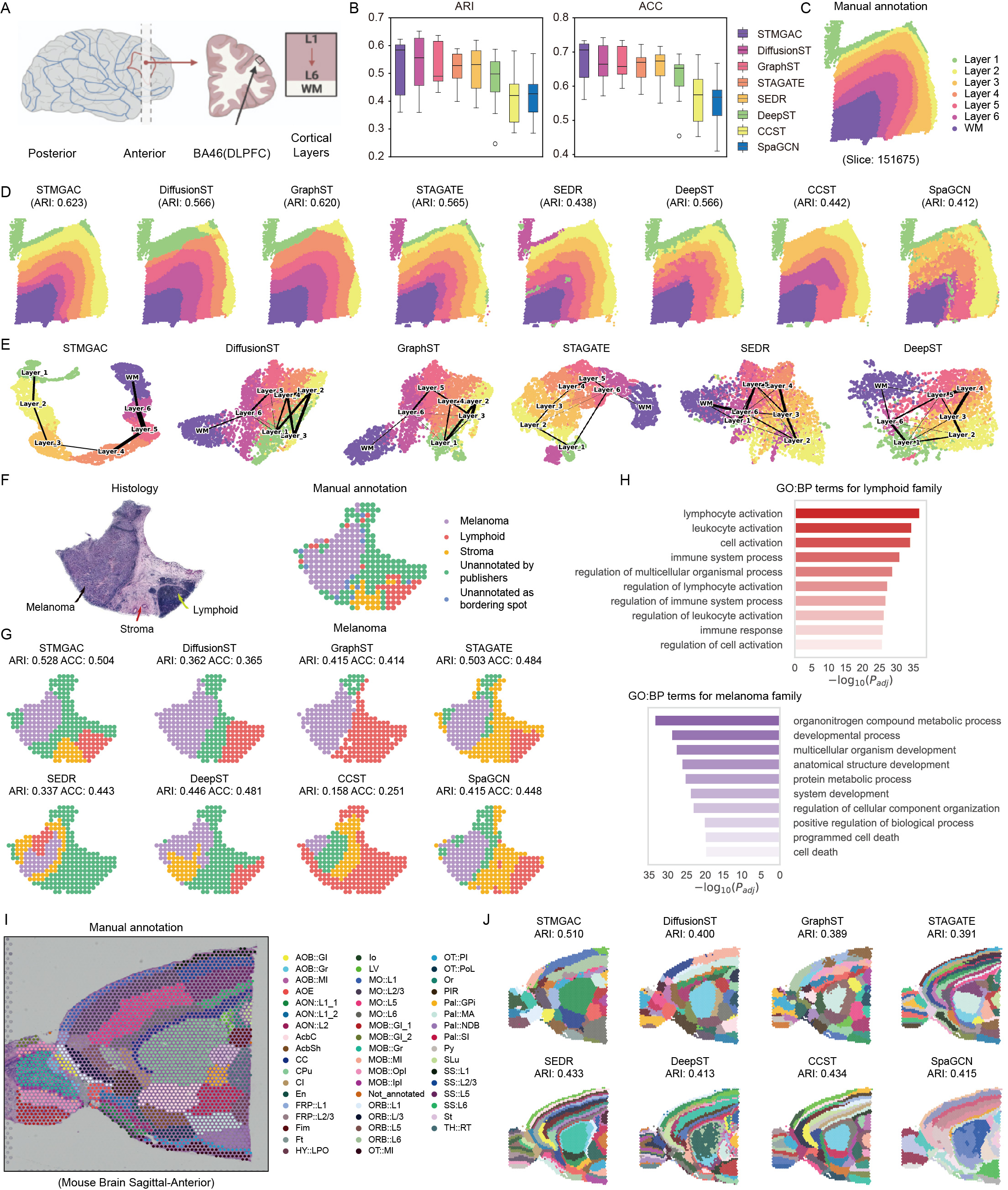}
\caption{STMGAC enables the identification of tissue structures. (A) Schematic of DLPFC data. (B) The evaluation of STMGAC and existing methods on the DLPFC dataset was conducted based on ARI and ACC. (C) Manual annotation for slice 151675. (D) We conduct a comparison of spatial domain identification on slice 151675. (E) The UMAP plot of slice embeddings colored cortical layers. (F) Manual annotation for HM. (G) Spatial domain identification on HM. (H) Here are the top 10 significant GO:BP terms for Cluster 2 (Lymphoid) and Cluster 3 (Melanoma). (I) Manual annotation for MBA. (J) Spatial domain identification on MBA.}
\label{fig2}
\end{figure*}

%%%
\begin{figure*}[!t]%
\centering
\includegraphics[width=1\textwidth]{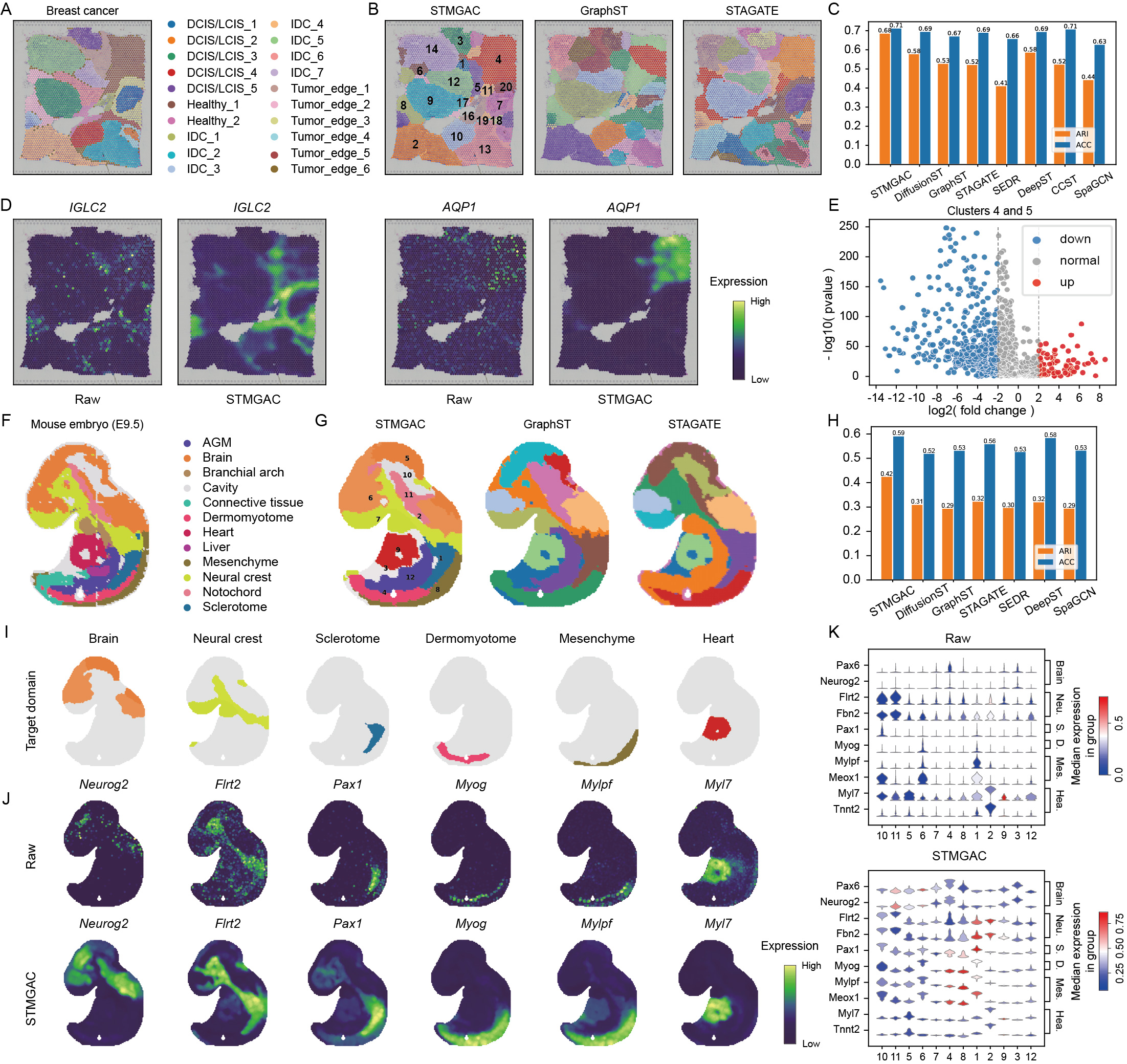}
\caption{STMGAC denoises gene expressions. (A) Manual annotation for BRCA. (B) Spatial domain identification using the STMGAC, GraphST, and STAGATE methods. (C) Bar chart of ARI and ACC metrics for various methods on BRCA. (D) Raw spatial expression and STMGAC denoised expression of IGLC2 and AQP1 genes. (E) Volcano plot of differentially expressed genes between Cluster 4 (health) and Cluster 5 (tumor edge). (F) Manual annotation for ME9.5. (G) Spatial domain identification. (H) Bar chart of ARI and ACC metrics for various methods on ME. (I) Spatial domains of major tissues identified by STMGAC. (J) Raw spatial expression and denoised expression by STMGAC. (K) Violin plots of the denoised expression of domain-specific marker genes by STMGAC.}
\label{fig3}
\end{figure*}

\section{EXPERIMENTS}\label{result}
\subsection{Dataset description}
In this study, we analyze the clustering performance of STMGAC on a range of SRT datasets from different platforms. First, three datasets from the 10x Visium platform include the human dorsolateral prefrontal cortex (DLPFC), human breast cancer (BRCA), and the anterior mouse brain tissue (MBA) datasets. The DLPFC dataset consists of three independent donors, with each donor having four slices. Each slice is manually annotated with 5 to 7 spatial domains  \cite{DLPFC}. Xu et al. manually annotated 20 different domains in the BRCA dataset \cite{SEDR}. The Allen Mouse Brain Reference Atlas manually annotates the MBA dataset into 51 known clusters and one unmarked region \cite{allen}. 

In addition, there is the human melanoma (HM) dataset from the ST platform, where Thrane et al. manually annotated three distinct regions \cite{HM}: melanoma, stroma, and lymphoid tissue, along with an additional unannotated region. Finally, there is the mouse embryo (ME) dataset from the Stereo-seq platform, representing the E9.5 developmental stage \cite{ME}, where Chen et al. manually annotated 12 tissue regions \cite{ME_annotation}. More information about the datasets can be found in \autoref{tab-1}.

\begin{table}[t]
\caption{The statistics of the datasets used in this study.}\label{tab-1}
\begin{adjustbox}{width=0.5\textwidth}
\begin{tabular}{l|c|c|c|c|c}
\hline
Datasets      & \#Slices  & \#Spots  & \#Genes   &  \#Domains  & Platform                  \\ \hline
DLPFC \cite{DLPFC}   & 12   & 3,460-4,789 & 33,538    & 5-7  & 10x Visium    \\
BRCA \cite{SEDR}        & 1  & 3,798 & 36,601     & 20  & 10x Visium       \\
MBA \cite{allen}        & 1  & 2,695 & 32,285     & 52  & 10x Visium       \\
HM \cite{HM}        & 1  & 293 & 16,601     & 4  & ST       \\
ME   \cite{ME}      & 1  & 5,913 & 22,385     & 12   & Stereo-seq     \\ \hline
\end{tabular}
\end{adjustbox}
\end{table}

\subsection{Baseline methods}
We selected several state-of-the-art methods that are representative of the following:
\begin{itemize}
    \item SpaGCN \cite{SpaGCN} integrates gene expression, spatial location, and histology in SRT data via graph convolution.
    
    \item CCST \cite{ccst} is developed based on the node embedding method DGI to achieve spatial domain clustering.

    \item DeepST \cite{DeepST} employs GCN as an encoder to reconstruct the input graph topology and capture the spot features.

    \item SEDR \cite{SEDR} utilizes a deep autoencoder network to learn gene expression while simultaneously incorporating the spatial information from the variation graph autoencoder.

    \item STAGATE \cite{STAGATE} leverages adaptive graph attention autoencoders to integrate spatial location and gene expression.

    \item GraphST \cite{graphst} is a graph self-supervised contrastive learning method that leverages spatial information and gene expression profiles.

    \item DiffusionST \cite{Diffusionst} employs a ZINB distribution and diffusion models to denoise and enhance SRT data.
\end{itemize}

%\subsection{Experimental setup}
\subsection{Implementation Details}
For STMGAC, we use a learning rate of 0.001 and a weight decay of 2e-4, optimized with Adam. The online and momentum encoders have linear layers of dimensions 64 and 32, followed by GCN layers with output dimensions of 64 and 16. The predictor dimension is 32, and the feature decoder reconstructs to the raw data space. The default masking rate is 0.5. For anchor pairs selection, the 10x Visium and Stereo-seq datasets use the top 50 nearest neighbors, while ST uses the top 30 nearest neighbors. Default parameters from the original papers were applied for all baselines, and experiments were conducted on an NVIDIA GeForce RTX 3090. 

\subsection{STMGAC enables the identification of tissue structures from SRT data}
\subsubsection{Applying STMGAC to the DLPFC dataset} 
We compared the clustering results of the STMGAC model with seven state-of-the-art representative algorithms on the DLPFC dataset. The bar chart in \autoref{fig2}B shows the ARI and ACC values of each method across 12 slices. As shown in the figure, STMGAC achieved the highest median ARI (0.584) and ACC (0.706), while the existing benchmark methods had median ACC values below 0.70. 

% Furthermore, we demonstrated the spatial domain identification results using slice 151675 as an example (\autoref{fig2}C and D). It can be observed that SpaGCN, SEDR, and DeepST have a large number of misclassified spots across different domains, with unclear boundaries. CCST could not distinguish between layer 5 and layer 6. DiffusionST and GraphST failed to detect layer 4. Only STMGAC identified all layers with very clear boundaries and no mixed spots. 
We showed spatial domain identification results for slice 151675 (\autoref{fig2}C and D). SpaGCN, SEDR, and DeepST had many misclassified spots and unclear boundaries. CCST could not distinguish layers 5 and 6, while DiffusionST and GraphST missed layer 4. Only STMGAC correctly identified all layers with clear boundaries and no mixed spots.

From the UMAP visualization results (\autoref{fig2}E), it was found that only the embeddings of STMGAC were consistent with the development trajectory of the layers, showing a linear pattern. It is worth noting that both DiffusionST and GraphST used label refinement techniques, which improved performance but did not substantially change the UMAP visualization trajectory of their embeddings. 

\subsubsection{Applying STMGAC to the HM dataset} 
% In the HM dataset (\autoref{fig2}F and G), the baseline methods failed to detect the stroma region, but STMGAC successfully identified it. STMGAC demonstrated superior clustering performance, while CCST performed the worst due to the dataset’s spatial resolution of 100 µm, which has fewer spots compared to 10x Visium. This led to over-smoothing by CCST on the sparse and low-density HM dataset.
In the HM dataset (\autoref{fig2}F and G), baseline methods missed the stroma region, whereas STMGAC successfully identified it. STMGAC showed superior clustering performance. CCST performed poorly on the sparse and low-density HM dataset, leading to excessive smoothing.

To explore gene expression differences, we analyzed differential expression in Cluster 2 (lymphoid family) and Cluster 4 (melanoma family) using the criteria $|\log2(\text{FoldChange})| \ge 2$ and $P < 0.05$. The identified genes play key roles in various physiological and pathological processes. We then performed gene enrichment analysis using the Gene Ontology: Biological Process (GO:BP) \cite{gobp} database (\autoref{fig2}H). For Cluster 2, the top 10 GO:BP terms were mainly related to cell morphology and immune response regulation. For Cluster 4, the terms focused on biological regulation and metabolic processes \cite{min2024dimensionality}.

% To further investigate gene expression differences, we conducted differential expression analysis on Cluster 2 (lymphoid family) and Cluster 4 (melanoma family) using the criteria $|\log2(\text{FoldChange})| \ge 2$ and $P value  < 0.05$. These differentially expressed genes are crucial in various physiological and pathological processes. We then performed gene enrichment analysis on the identified genes within these clusters using the Gene Ontology: Biological Process (GO:BP) \cite{gobp} database (\autoref{fig2}H). Cluster 2 was identified as the lymphoid family, with the top 10 GO:BP terms primarily related to changes in cell morphology and behavior, as well as immune response regulation. Cluster 4 corresponded to the melanoma family, with terms predominantly associated with biological regulation and metabolic processes \cite{min2024dimensionality}.

\subsubsection{Applying STMGAC to the MBA dataset}
We analyzed the MBA dataset (\autoref{fig2}I), which presents a more complex tissue structure compared to the DLPFC dataset, posing greater challenges. STAGATE incorrectly divided the central nucleus (CPu) into multiple regions, while DeepST and SpaGCN showed significant spot mixing, resulting in unclear cluster boundaries. STMGAC demonstrated an improved ability to identify spatial domains in complex tissue structures (\autoref{fig2}J).

\subsection{STMGAC denoises gene expressions for better characterizing spatial expression patterns}

\subsubsection{Applying STMGAC to the BRCA dataset} We used STMGAC to reduce noise in the BRCA dataset and better display the spatial patterns of genes. First, we identified differentially expressed genes in Cluster 4 (health) and Cluster 5 (tumor edge), with 564 and 84 genes respectively, and plotted their differential expression using volcano plots (\autoref{fig3}E). For example, after denoising, the \emph{IGLC2} and \emph{AQP1} genes showed differential expression in the tumor edge and health domains, respectively, whereas their raw spatial expression was completely disordered (\autoref{fig3}D).

\subsubsection{Applying STMGAC to the ME dataset} We evaluated the denoising capability of STMGAC on the ME dataset by comparing the expression of key marker genes across six tissue regions between the raw data and the STMGAC-denoised data. The results demonstrated significant spatial enrichment of gene expression in these tissue regions post-denoising (\autoref{fig3}I and J). For instance, the \emph{Neurog2} gene exhibited differential expression in the brain region after denoising, which aligns with its known function. Similarly, the \emph{Myl7} gene, which is involved in calcium ion binding activity, was primarily expressed in the heart and vascular system, consistent with previous findings \cite{ME, ME_annotation}. Additionally, violin plots comparing the raw and STMGAC-denoised data for 10 marker genes indicated that STMGAC significantly enhanced the spatial expression patterns of these genes (\autoref{fig3}K).

\subsection{Ablation studies}

\begin{table}[]
\caption{Ablation study on the impact of different components}\label{ablation1}
\begin{adjustbox}{width=0.49\textwidth}
\begin{tabular}{l|cccccc}
\toprule%
\multirow{2}{*}{Method}   & \multicolumn{2}{c}{BRCA} & \multicolumn{2}{c}{HM} & \multicolumn{2}{c}{ME} \\ 
\addlinespace[-0.8ex] % 手动减少行间距
\cmidrule(lr){2-3}\cmidrule(lr){4-5}\cmidrule(lr){6-7}
\addlinespace[-0.8ex] % 手动减少行间距
                          & ARI         & ACC        & ARI        & ACC       & ARI        & ACC       \\ \midrule
w/o matching loss         & 0.6298      & 0.6938     & 0.4561     & 0.4587    & 0.3465     & 0.5709    \\
w/o GCN predictor         & 0.6219      & 0.6891     & 0.4307     & 0.4511    & 0.3592     & 0.5710    \\
w/o EMA (Shared encoder)  & 0.6216      & 0.6931     & 0.5047     & 0.4907    & 0.3657     & 0.5825    \\
w/o triplet loss          & 0.5675      & 0.6851     & 0.4554     & 0.4248    & 0.3419     & 0.5381    \\
w/o negative nodes        & 0.5911      & 0.6701     & 0.4034     & 0.3857    & 0.3303     & 0.5533    \\
w/o local positive nodes  & 0.6665      & 0.6970     & 0.5116     & \textbf{0.5051}    & 0.4016     & 0.5670    \\
w/o global positive nodes & 0.6334      & 0.6998     & 0.4872     & 0.4566    & 0.3828     & \textbf{0.5962}    \\
STMGAC                    & \textbf{0.6842}      & \textbf{0.7108}     & \textbf{0.5278}     & 0.5043    & \textbf{0.4234}     & 0.5898    \\ \bottomrule
\end{tabular}
\end{adjustbox}
\end{table}

\begin{table}[]
\caption{Ablation study on the impact of different loss functions}\label{ablation2}
\begin{adjustbox}{width=0.49\textwidth}
\begin{tabular}{ccc|cccccc}
\toprule%
\multicolumn{3}{c|}{Loss Function} & \multicolumn{2}{c}{BRCA} & \multicolumn{2}{c}{HM} & \multicolumn{2}{c}{ME} \\
\addlinespace[-0.8ex] % 手动减少行间距
\cmidrule(lr){1-3}\cmidrule(lr){4-5}\cmidrule(lr){6-7}\cmidrule(lr){8-9}
\addlinespace[-0.8ex] % 手动减少行间距
$\mathcal{L}_R$        & $\mathcal{L}_M$        & $\mathcal{L}_T$        & ARI         & ACC        & ARI        & ACC       & ARI        & ACC       \\ \midrule
MSE       & MSE       & CON       & 0.5101      & 0.6469     & 0.5162     & 0.4462    & 0.3441     & 0.5455    \\
MSE       & MSE       & TRI       & 0.6749      & 0.6895     & \textbf{0.5428}     & \textbf{0.5073}    & 0.3987     & 0.5271    \\
MSE       & SCE       & CON       & 0.5994      & 0.6795     & 0.5192     & 0.4729    & 0.3393     & 0.5345    \\
MSE       & SCE       & TRI       & 0.6735      & 0.6821     & 0.5026     & 0.4776    & 0.3391     & 0.5452    \\
SCE       & MSE       & CON       & 0.5615      & 0.6685     & 0.3888     & 0.3976    & 0.3607     & 0.5669    \\
SCE       & MSE       & TRI       & \textbf{0.6842}      & \textbf{0.7108}     & 0.5278     & 0.5043    & \textbf{0.4234}     & \textbf{0.5898}    \\
SCE       & SCE       & CON       & 0.6122      & 0.6867     & 0.4485     & 0.4686    & 0.4015     & 0.5737    \\
SCE       & SCE       & TRI       & 0.6128      & 0.6927     & 0.5204     & 0.5060    & 0.3940     & 0.5847    \\ \bottomrule

\end{tabular}
\end{adjustbox}
\end{table}

\subsubsection{Contributions of different components} To investigate the contributions of different components to STMGAC, we designed several variants of STMGAC and explored their impact on model accuracy across three datasets from different platforms (BRCA, HM, and ME datasets) (\autoref{ablation1}).

For the latent space supervision signal process, we designed three variants: w/o matching loss (without latent space supervision); w/o GCN predictor (using MLP to predict latent representation directly without the remask technique); w/o EMA (sharing weights with the online encoder while the momentum encoder is detached from gradient back-propagation). For the anchor selection method, we designed four variants: w/o triplet loss (completely missing this component); w/o negative nodes (using only positive nodes); w/o local positive nodes; w/o global positive nodes.

The results demonstrated that STMGAC achieved the best performance across multiple datasets, only performing slightly worse on the sparse and low-density HM dataset. This slight underperformance is due to the small number of spots, which requires higher accuracy in anchor selection.

\subsubsection{Impact of different loss functions} We investigated the impact of reconstruction loss in the raw space, representation matching loss in the latent space, and contrastive loss. Both \(\mathcal{L}_r\) and \(\mathcal{L}_m\) used Mean Squared Error loss (MSE) and Scaled Cosine Error loss (SCE), while \(\mathcal{L}_t\) used triplet loss (TRI) and contrastive discrimination loss (CON). As a result, using SCE in the raw space effectively learned features, while using MSE in the latent space for mask matching made a significant contribution. TRI greatly enhanced clustering effects (\autoref{ablation2}).

\section{CONCLUSIONS AND DISCUSSION}
In this paper, we propose a Masked Graph Autoencoder with Contrastive Augmentation (STMGAC) method for clustering and gene denoising analysis of SRT data. Previous graph masking methods did not focus on the supervision signals in the latent space, making it difficult for autoencoders to reconstruct raw data accurately. To address this, we designed the raw gene expression data as a masked matrix and a visible spot matrix, utilizing an EMA mechanism to provide persistent and reliable supervision signals for the model's representation in the latent space. Furthermore, in the latent space, we select positive and negative anchor pairs based on the correlation between representations, local adjacency relationships, and global information, bringing similar instances closer together and pushing dissimilar instances further apart, thereby effectively enhancing spatial domain recognition capabilities.

We analyzed the performance of STMGAC on dataset from different platforms and achieved results superior to the existing seven baseline methods. Additionally, ablation studies fully explored the contributions of each component to STMGAC.

\section*{Acknowledgment}
The work was supported in part by the National Natural Science Foundation of China (62262069), in part by the Yunnan Fundamental Research Projects under Grant (202201AT070469) and the Yunnan Talent Development Program - Youth Talent Project. 

\balance
\small
\bibliography{references.bib} %.bib文件名字
\bibliographystyle{IEEEtran}  %.bst模板	
\end{document}